\documentclass[preprint,12pt,number]{elsarticle}

\usepackage{hyperref}
\usepackage{subfig}
\usepackage{amsfonts}
\usepackage{enumitem}

\usepackage{multirow}
\usepackage{multicol}
\usepackage{float}

\usepackage{amsmath}
\usepackage{algorithm}
\usepackage{algpseudocode}

\journal{Elsevier}

\begin{document}

\begin{frontmatter}

\title{A Membrane Computing Approach to \\the Generalized Nash Equilibrium}

\author[a]{Alejandro Luque-Cerpa\corref{cor1}}
\ead{luque@chalmers.se}
\author[b]{Miguel A. Gutiérrez-Naranjo}
\ead{magutier@us.es}

\cortext[cor1]{Corresponding author}
\affiliation[a]{organization={Department of Computer Science and Engineering, Chalmers University of Technology},
country={Sweden}}
\affiliation[b]{organization={Department of Computer Sciences and Artificial Intelligence, University of Seville},
country={Spain}}

\begin{abstract}
In Evolutionary Game Theory (EGT), a population reaches a Nash equilibrium when none of the agents can improve its objective by solely changing its strategy on its own. Roughly speaking, this equilibrium is a protection against betrayal. Generalized Nash Equilibrium (GNE) is a more complex version of this idea with important implications in real-life problems in economics, wireless communication, the electricity market, or engineering among other areas. In this paper, we propose a first approach to GNE with Membrane Computing techniques and show how GNE problems can be modeled with P systems, bridging both areas and opening a door for a flow of problems and solutions in both directions. 
\end{abstract}


\end{frontmatter}

\section{Introduction}
EGT studies the evolution of a population of agents that interact with each other and get a payoff in each interaction \cite{EGT}. The obtained payoff depends on the chosen strategies of the agent which participate in the interaction. Each agent selects only one strategy at a time, but this choice can be modified for the next steps.
The driving principle in this situation is that individuals tend to choose strategies that provide them with higher payoffs. In this context, a Nash equilibrium is reached when no agent can increase its payoff by changing its strategy while other agents maintain their current ones \cite{Nash1951}. 

In a Nash equilibrium problem, all the agents compete among them 
to maximize their payoffs, and each agent can freely choose its strategy. The generalized Nash equilibrium problem (GNEP) is a variant of the Nash problem introduced in 1952 by G. Debreu \cite{Debreu1952}. In a GNEP, the strategy set of each player may also depend on the other players' strategies. This GNEP models a large number of real-life situations, such as power allocation in a telecommunication system, environmental pollution control, or energy market model (for a detailed survey, see, e.g., \cite{DBLP:journals/4or/FacchineiK07}).

In this paper, we propose to study the GNEP in the framework of Membrane Computing. Membrane Computing \cite{paunbook,HandbookMC10} is a well-known area in Computer Science that takes inspiration from the biochemical reactions inside the vesicles of living cells. P systems, the so-called Membrane Computing devices, have been successfully considered to model many dynamic processes in real-life problems \cite{scavengers, chamois, butterfly}. From the initial definition of P systems, \cite{Paun98}, the devices of Membrane Computing, many variants have been explored, by adding new features to the initial model (see, e.g., \cite{10.1145/3431234} for a recent survey). Recently, Probabilistic P systems \cite{DBLP:journals/nc/CardonaCMPPPS11}, a kind of P system designed to deal with probability distributions in the application of rules, was considered to model the spread of behaviors in structured populations in the framework of EGT \cite{DBLP:journals/isci/Garcia-Victoria22}. In this paper, we study the GNEP by considering transition P systems with membrane polarization.

The paper is organized as follows: Section 2 introduces population games under the Brown-von Neumann-Nash (BNN) dynamics, that will be used as the framework to define our P system. Section 3 establishes some background on P systems and specifies the type of P system we use: transition P systems with membrane polarization. In Section 4, we describe the design of our P system and analyze its computation, proving that it is linear with respect to time, and does not depend on the number of players or strategies. We also present an experiment to illustrate its functioning. Finally, some conclusions and hints for future work are presented.

\section{Population Games under BNN Dynamics}\label{section:Game}
In population games, each agent is endowed with a revision protocol, which provides conditional switch rates between strategies according to their associated payoffs \cite{populationgames}. These rates allow the agents to change their strategies over time. When the number of agents is large enough, this process can be described by differential equations, referred to as the evolutionary dynamics model (EDM). There are multiple EDMs, but we focus on a specific EDM known as BNN dynamics \cite{BrownNeumann+1951+73+80}.  

Next, some basics of this model are recalled (see \cite{DBLP:conf/eucc/Martinez-Piazuelo22} for details). Let us consider a society of agents divided into $N \in \mathbb{Z}_{\geq 1}$ disjoint populations indexed by $\mathcal{P} = \{1, 2, ..., N\}$. Each population $k \in \mathcal{P}$ is comprised of a constant mass of decision-making agents $m^k \in \mathbb{R}_{>0}$. The set of strategies of each agent in population $k \in \mathcal{P}$ is $\mathcal{S}^k \subset \mathbb{Z}$ with $n^k = |\mathcal{S}^k| \geq 2$. The amount of agents selecting strategy $i \in \mathcal{S}^k$ at population $k$ is denoted as $x_i^k \in \mathbb{R}_{\geq 0}$. Similarly, the proportion of agents selecting strategy $i \in \mathcal{S}^k$ at population $k$ is denoted as $z_i^k = x_i^k / m^k$. Furthermore, $x^k = $col$(x_{i_1}^k, ..., x_{i_{n^k}}^k)$ and $z^k = $col$(z_{i_1}^k, ..., z_{i_{n^k}}^k)$ denote the strategic distributions of population $k$, and $x = $col$(x^1, x^2, ..., x^N)$, $z = $col$(z^1, z^2, ..., z^N)$. Knowing the values $(i,k)$ of the element $x_i^k$, the position $l$ of the element $x_l$ in the vector $x$ can be computed (and vice versa). 
Let $t \in \mathbb{Z}_{\geq 0}$ be the discrete-time index; $x(t)$ the value of $x$ at time $t$; $z(t)$ the value of $z$ at time $t$ and $p_i^k(t) \in \mathbb{R}$ the payoff received by the agents selecting strategy $i \in \mathcal{S}^k$ at population $k \in \mathcal{P}$.

Following the revision protocol introduced in \cite{DBLP:conf/eucc/Martinez-Piazuelo22}, the equations that define the EDM describing the evolution of $x(t)$ over time are:
\begin{align} 
\dot{x}_i^k(t) = m^k [\hat{p}_i^k(t)]_+  - x_i^k(t) \sum\limits_{j \in \mathcal{S}^k} [\hat{p}_j^k(t)]_+ \label{eq1}  \\ 
\hat{p}_j^k(t) = p_j^k(t) - \frac{1}{m^k} \sum\limits_{l \in \mathcal{S}^k} x_l^k(t) \cdot p_l^k(t) \label{eq2}
\end{align}
where $[\cdot]_+ = \max(\cdot,0)$. This EDM is known as the BNN dynamics \cite{BrownNeumann+1951+73+80}. A payoff dynamics model (PDM) that describes the evolution of $p(t)$ is also introduced, defined by:
$$ \dot{\mu}(t) = Ax(t) - b$$
$$ p(t) = f(x(t)) - A^\top \mu(t) $$

Since the importance of this system lies in updating the payoff signal $p(t)$ and having a closed-loop configuration between $p(t)$ and $x(t)$, a simplified version of this system is considered:
\begin{align}  p(t) = f(x(t)) \label{eq3}\end{align}

To define our P system, a specific function $f$ must be considered. Taking this into account, a specific EDM must be selected as a framework. If another EDM were to be selected, only the first stage of the P system would have to be modified, while the rest would remain unchanged. Because of this, an example of the Energy Market Game \cite{DBLP:conf/eucc/Martinez-Piazuelo22} is considered as the framework, where $N \in \mathbb{Z}_{\geq 1}$ players compete to purchase energy over a time horizon of $T \in \mathbb{Z}_{\geq 1}$ time slots. In this game, let $C^k \in \mathbb{R}^{T \times n^k}$ be a matrix such that each column of $C^k$ has exactly one element equal to 1 and the rest equal to 0, each row of $C^k$ has at most one element equal to 1 and the $j$-th element of the $i$-th column of $C^k$ is 1 iff player $k$ competes in time slot $j \leq T$. Let $C = [C^1, C^2, \dots, C^N] \in \mathbb{R}^{T \times n}$ be the concatenation of the $C^k$ matrices of all players. Then $Cx$ corresponds to the collective energy demand for all time slots. Let $J : \mathbb{R}^n \rightarrow \mathbb{R}^T$ be the pricing function given by $J(x) = DCx + \bar{J}$, where $D \in \mathbb{R}^{T \times T}_{\geq 0}$ is diagonal and $\bar{J} \in \mathbb{R}^T_{\geq 0}$, and let $Q^k : \mathbb{R}_{\geq 0}^{n^k} \rightarrow \mathbb{R}$ be the individual cost of each player $k \in \mathcal{P}$, given by $Q^k(x^k) = \sum\limits_{i \in S^k} \big( (\alpha_i^k / 2) (x_i^k)^2 + \beta_i^k \cdot x_i^k \big)$, where $\alpha_i^k \in \mathbb{R}_{\geq 0}$ and $\beta_i^k \in \mathbb{R}_{\geq 0}$. To define the payoff over $z_i^k(t)$, the following transformation is performed over Eq. \ref{eq3}: 
    \begin{align}
        p(t) = f(x(t)) = f(M \cdot z(t)) = -S \cdot M \cdot z(t) - C^\top \bar{J} - \alpha \odot (M  \cdot z(t)) - \beta \label{eq4}
    \end{align}  
where 
 \begin{itemize}
\item $M = $diag$(m^1 \textbf{1}_{n^1}, m^2 \textbf{1}_{n^2}, \dots, m^N \textbf{1}_{n^N})$, 
\item $S = $diag$(C^{1 \top} D C^1, \dots, C^{N \top} D C^N) + R^\top R$, and 
\item $R = [\sqrt{D}C^1,\sqrt{D}C^2, \dots, \sqrt{D}C^N]$.
\end{itemize}

The system equations Eq. \ref{eq1}, Eq. \ref{eq2} also change for $z_i^k(t)$: 
    \begin{align}
        \dot{z}_i^k(t) &= \frac{\dot{x}_i^k(t)}{m^k} = [\hat{p}_i^k(t)]_+ - z_i^k(t)  \cdot \sum\limits_{j \in \mathcal{S}^k} [\hat{p}_j^k(t)]_+ \label{eq5} \\
        \hat{p}_j^k(t) &= p_j^k(t) -  \sum\limits_{l \in \mathcal{S}^k} z_l^k(t) \cdot p_l^k(t) \label{eq6}
    \end{align} 

\section{Transition P systems with membrane polarization}
After the development of the first model of membrane computing by Gh. P\v{a}un in 1998 \cite{PAUN2000108}, many variations have been presented. In this work, we use a combination of two proposed variants. The P system designed is a transition P system \cite{PAUN2000108} with active membranes \cite{paunactive} without division rules, i.e., a transition P system with membrane polarization. Such a system is a construct:
$$ \Pi = \langle \Gamma, H, EC, \mu, \{w_h\}_{h \in H}, (\mathcal{R}, \rho) \rangle $$
where:
\begin{enumerate}
    \item $\Gamma$ is the alphabet of \textit{objects};
    \item $H$ is a finite set of \textit{labels} for membranes;
    \item $EC$ is a finite set of electrical charges;
    \item $\mu$ is a \textit{membrane structure}, labeled with elements of \textit{H} where different membranes have different labels and each membrane has a polarization in $EC$;
    \item $w_h$ are strings over $\Gamma$, describing the \textit{multisets of objects} placed in the regions of $\mu$;
    \item $\mathcal{R}$ is a finite set of \textit{evolution rules} of the form $r \equiv u [v]_i^\alpha \rightarrow u' [v']_i^\beta$, where $u,u',v,v'$ are multisets over $\Gamma$, $i$ is a membrane label, and $\alpha, \beta \in EC$. \\ $\rho$ is a partial order relation over $\mathcal{R}$, called a \textit{priority} relation. Given two rules $r, r'$, we represent that $r$ has higher priority than $r'$ by $\rho_{r} > \rho_{r'}$. Priority indicates what rule should be applied if both can be applied. For each configuration, the rules are applied in a parallel and maximal way, this is, all rules that can be applied are applied to generate the next configuration.
\end{enumerate}

As in \cite{DBLP:journals/isci/Garcia-Victoria22}, the semantics of the P system follow the next principles:

\begin{enumerate}[label=(I\arabic*)]
    \item When an object crosses a membrane, its polarization may change. Rules can only be applied if the polarization is appropriate.
    \item If a rule can be applied inside a membrane and, in the same step, a rule that changes the polarization of a membrane can also be applied, both rules are applied. This means that the change of polarization is performed \textit{after} all other evolution rules are applied. 
\end{enumerate}

\section{Design and functioning of the P System \label{section:NE}}
Let us consider the EDM-PDM system with the parameters introduced in Section \ref{section:Game}. In this section, a P system that computes approximations of GNE under the BNN dynamics is described. The computation can be summarized in a loop of five stages, represented in Algorithm \ref{alg:psystem}. Stages 1 and 2 are used to compute $\hat{p}(t)$ using Eq. \ref{eq6}, Stages 3 and 4 are used to compute $\dot{z}(t)$ using Eq. \ref{eq5}, and Stage 5 is used to update the value of $z(t)$.

The fundamental idea behind the system is to discretize the EDM-PDM system by rounding everything to two decimals and to evolve objects representing $z(t)$ to compute GNE. In this context, an object representing $z$ would represent 1\% of the agents. The system can easily be modified to round to other quantities. To round, we perform round$(x) = \lfloor 100 \cdot x \rfloor$. To evolve the system, we use Euler's method \cite{EulerMethodBook}. 

In Algorithm \ref{alg:psystem}, other stop conditions can be easily defined, for example, comparing the $z(t)$ values of one iteration with those of the previous one (in constant time) and stopping if no difference is found, but more rules would be necessary. For the sake of simplicity, our stop condition is to limit the number of iterations in the loop.

To perform multiplications, a P system that replicates the Russian peasant multiplication algorithm is defined in \ref{appendix:russian}. The complete P system is defined in \ref{appendix:DefinitionPSystem}. 

\begin{algorithm}
\caption{General overview of the P system computation}\label{alg:psystem}
\begin{algorithmic}
\Require $L \geq 0$, $t \geq 0$, $s=0$
\While{$s \leq L$}
\State 1. Computation of payoff $p(t)$ for current iteration
\State 2. Computation of sums $\sum\limits_{j \in S^k} z_j^k(t) \cdot p_j^k(t)$
\State 3. Computation of $[ \hat{p}_i^k ]_+$ and $[\sum\limits_{j \in S^k} \hat{p}_j^k ]_+$
\State 4. Computation of $\dot{z}_i^k(t)$
\State 5. Update of $z(t)$, coordination for next iteration and results output
\State $s$ += $1$
\EndWhile
\end{algorithmic}
\end{algorithm}

\subsection{Overview of the computation}
In this subsection, an analysis of the computation for each stage is provided. The result of the computation analysis is that for each time step $t$, the number of transition steps is bounded by a constant. This means that the complexity of the global computation only grows linearly with $t$, and it does not depend on the number of players or the strategies involved. Only the membranes and objects involved in each stage are considered during the analysis. \newline
 
    \textbf{Stage 1: Computation of payoff $p(t)$} \\ In this stage, the payoff of the current iteration is computed. To do this, it is necessary to consider the function $f$ involved in Eq. \ref{eq4}.  
    The computation starts with the initial configuration. If we only consider the membranes involved in this stage, we have the following configuration: 
    $$ \mathbb{C}_0 = [[[\langle k,i,l \rangle^{z^k_i}]_{S_{i,k}}^0 ]_k^0 [y_0]_P^0]_0^0 $$
    After 8 transitions steps, each value $p_l(t)$ of $p(t)$ is computed:
    $$ \mathbb{C}_8 = [[ \langle a,k,i,l \rangle^{p_l(t)} ]_k^0  [y_6 ]_P^- ]_0^0 $$
    Some rules in Stage 2 will transform the objects $\langle a,k,i,l \rangle$ that are in the membranes $[ \ ]_k$, but for now, we are interested in seeing when Stage 1 will finish. Two iterations later:
    $$ \mathbb{C}_{10} = [[ \dots mult_0^{|S^k|} ]_k^- \ [ \ ]_P^0 ]_0^0 $$
    
    \textbf{Stage 2: Computation of the sums $\sum\limits_{j \in S^k} z_j^k(t) p_j^k(t)$} \\ Stage 2 starts with the last configuration of Stage 1: $\mathbb{C}_{10}$. After $\mathbb{C}_{2}$ and $\mathbb{C}_{8}$, there are some elements $\langle Prod,k,i,l \rangle$ and $\langle a,k,i,l \rangle$ in $[ \ ]_k$, and these membranes $k$ change their polarization from $0$ to $-$ in $\mathbb{C}_{10}$, initiating Stage 2. After only two transition steps:
    $$ \mathbb{C}_{12} = [[ [ b^{p_l(t)} [ a^{z^k_i} k_1 ]_{M1}^0 [ \ ]_{M2}^0 ]_{MULT_{i,k}}^0 \  [ neg_i^{p_l(t)}  ]^0_{UPD_{i,k}} \ [ \ ]^0_{ACUM_{k}} rem^{|S^k|} ]_k^-]_0^0 $$
    In this configuration, $p_l=p_i^k$, so $z_i^k\cdot p_i^k$ is computed in each membrane $MULT_{i,k}$. Because the round function multiplies by 100, the (rounded) result of the multiplication will be $100 \cdot z_i^k \cdot p_i^k$. Furthermore, each product is computed after (at most) 43 iterations, as proven in \ref{appendix:russian}. Some objects $d$ may have been introduced as objects $pos$ into membranes $ACUM_k$, so after (at most) 44 iterations, and ignoring membranes $MULT_{i,k}$:
    $$ \mathbb{C}_{\leq 56} = [[[ neg_i^{p_l(t)}  ]^0_{UPD_{i,k}} \ [ pos^{100 \cdot z_i^k \cdot p_i^k} y_{2,0} ]^+_{ACUM_{k}} ]_k^-]_0^0 $$
    After three transition steps:
    $$ \mathbb{C}_{\leq 59} = [[[ neg_i^{p_l(t)}  ]^0_{UPD_{i,k}} \ [ \ ]^0_{ACUM_{k}} \ pos^{\sum\limits_{i \in S^k} z_i^k \cdot p_i^k} \ y_{3,0} \ rem]_k^0]_0^0 $$

    \textbf{Stage 3: Computation of $[ \hat{p}_i^k ]_+$ and $\sum\limits_{j \in S^k} [\hat{p}_j^k ]_+$} \\ This stage starts with the last configuration of Stage 2: $\mathbb{C}_{\leq 59}$. The change of polarization of membranes $[ \ ]_k$ from $-$ to $0$ and the presence of objects $y_{3,0}$ initiate Stage 3. Following the rules of Stage 3, we can see that after 7 steps we have the configuration:
    $$ \mathbb{C}_{\leq 66} = [[[   ]^0_{UPD_{i,k}} q^{\sum\limits_{j \in S^k}[\hat{p}_j^k]_+}\ q_i^{[\hat{p}_i^k]_+} \  y_{3,7,i}]_k^0]_0^0 $$

    \textbf{Stage 4: Computation of $\dot{z}_i^k(t)$} \\  Stage 4 starts with the last configuration of Stage 3: $\mathbb{C}_{\leq 66}$. After objects $\langle Prod2,k,i,l \rangle$ were created in computation $\mathbb{C}_{11}$, they are processed and introduced in membrane $MULT2_{i,k}$ in computation $\mathbb{C}_{12}$, but the product is not initiated until all objects $y_{3,7,i}$ are generated:
    $$ \mathbb{C}_{\leq 66} = [[ \ [ \ ]_{M1'}^0 [ \ ]_{M2'}^0  prod^{z^k_i} ]^0_{MULT2_{i,k}} [ \ ]^0_{S_{i,k}}  q^{\sum\limits_{j \in S^k}[\hat{p}_j^k]_+}\ q_i^{[\hat{p}_i^k]_+} \  y_{3,7,i}]_k^0]_0^0 $$
    Four iterations later, the multiplication in $MULT2_{i,k}$ can start:
    $$ \mathbb{C}_{\leq 70} = [[ \ [ a^{z^k_i} k_1 ]_{M1'}^0 [ \ ]_{M2'}^0  b^{\sum\limits_{j \in S^k}[\hat{p}_j^k]_+} ]^0_{MULT2_{i,k}} [ \ ]^0_{S_{i,k}} rem  \ q_i^{[\hat{p}_i^k]_+} ]_k^0]_0^0 $$
    After at most 43 iterations, as proven in \ref{appendix:russian}:
    $$ \mathbb{C}_{\leq 113} = [ [ \ ]^0_{S_{i,k}} d_i^{100 \cdot z_i^k \cdot \sum\limits_{j \in S^k}[\hat{p}_j^k]_+ } \ f_1^{|S^k|} \ q_i^{[\hat{p}_i^k]_+} ]_k^0]_0^0 $$
    After four more iterations:
    $$ \mathbb{C}_{\leq 117} = [ [ y_{4,3} \ zneg^{z_i^k \cdot \sum\limits_{j \in S^k}[\hat{p}_j^k]_+ } s_1^{[\hat{p}_i^k]_+}]^+_{S_{i,k}}  ]_k^0]_0^0 $$
    Now, depending on the result, the objects generated will be different. We represent by $zvar(p||n)$ the possibility of having objects $zvarp$ or $zvarn$. Considering that $\dot{z}_i^k = [\hat{p}_i^k]_+ - z_i^k \cdot \sum\limits_{j \in S^k}[\hat{p}_j^k]_+$, we have:
    $$ \mathbb{C}_{\leq 118} = [ [ y_{5,0} zvar(p||n)^{\dot{z}_i^k} ]^-_{S_{i,k}}  ]_k^0]_0^0 $$

    \textbf{Stage 5: Update of $z(t)$, coordination for next iteration and results output} \\ This stage starts with the last configuration of Stage 4: $\mathbb{C}_{\leq 118}$. In the initial multiset, there were some copies of $\langle AUX, 0 \rangle$ that have not been processed yet. Since configuration $\mathbb{C}_{1}$, there are object $c$ in membranes $S_{i,k}$ that have not been processed because the polarization has changed from $0$ to $-$ in the last configuration for the first time:
    $$ \mathbb{C}_{\leq 118} = [ [ [ \langle AUX, 0 \rangle ]^0_{RES_{i,k}} y_{5,0} \ zvar(p||n)^{\dot{z}_i^k} \ c^{z_i^k} ]^-_{S_{i,k}}  ]_k^0]_0^0 $$

    To update $z(t)$, we use Euler's method with $z(t+0.01) = z(t) + 0.01 \cdot \dot{z}(t)$. Depending on the objects existing at the end of Stage 4, there are three possible cases. Let $m_i = z_i^k + 0.01 \cdot \dot{z}_i^k$. Then:

    \begin{itemize}
        \item \textbf{Case 1}: If $0 \leq m_i < 100$, then three transition steps later:
        $$ \mathbb{C}_{\leq 121} = [ [ [ \langle AUX, 0 \rangle ]^0_{RES_{i,k}} y_{5,3} \ comp^{100 - m_i} \ p_1^{m_i} ]^0_{S_{i,k}} \ y_{5,3,i_1} \dots y_{5,3,i_{|S^k|}}  ]_k^0]_0^0 $$
        And finally:
        $$ \mathbb{C}_{\leq 122} = [ [ [ \langle AUX, 0 \rangle ]^0_{RES_{i,k}} y_{5,4}  ]^0_{S_{i,k}} \ compw_i^{100 - m_i} \ w_i^{m_i} \ y_{5,4} ]_k^0]_0^0 $$

        \item \textbf{Case 2}: If $m_i < 0$, then three transition steps later:
        $$ \mathbb{C}_{\leq 121} = [ [ [ \langle AUX, 0 \rangle ]^0_{RES_{i,k}} y_{5,3} \ comp^{100} \ n^{m_i} ]^0_{S_{i,k}} \ y_{5,3,i_1} \dots y_{5,3,i_{|S^k|}}  ]_k^0]_0^0 $$
        And finally:
        $$ \mathbb{C}_{\leq 122} = [ [ [ \langle AUX, 0 \rangle ]^0_{RES_{i,k}} y_{5,4}  ]^0_{S_{i,k}} \ compw_i^{100} \ n^{m_i} \ y_{5,4} ]_k^0]_0^0 $$

        \item \textbf{Case 3}: If $m_i \geq 100$, then three transition steps later:
        $$ \mathbb{C}_{\leq 121} = [ [ [ \langle AUX, 0 \rangle ]^0_{RES_{i,k}} y_{5,3} \ p^{m_i - 100} ]^0_{S_{i,k}} \ y_{5,3,i_1} \dots y_{5,3,i_{|S^k|}} w_i^{100}  ]_k^0]_0^0 $$
        Notice that, in this stage, objects $y_{5,3,i}$ are always created in the same configuration in each of the three cases. Finally:
        $$ \mathbb{C}_{\leq 122} = [ [ [ \langle AUX, 0 \rangle ]^0_{RES_{i,k}} y_{5,4}  ]^0_{S_{i,k}} \ w_i^{100} \ p^{m_i-100} \ y_{5,4} ]_k^0]_0^0 $$
    \end{itemize}

    Regardless of the result of each case, objects $w_i$ represent the new $z_i^k$, objects $compw_i$ represent the difference between $z_i^k$ and $100$, and objects $p$ and $n$ represent positive and negative overflow respectively. If both overflows exist, they will cancel each other for the next configuration. These objects are introduced because, while using Euler's method, there is nothing that assures that $0 \leq z_i^k(t) \leq 100$ $\forall t \geq 1$. The next configuration is then:
    $$ \mathbb{C}_{\leq 123} = [ [ [ \langle AUX, 0 \rangle ]^0_{RES_{i,k}} y_{5,5}  ]^0_{S_{i,k}} \ err^u \ w_i^{\hat{z}_i^k} \ compw_i^{100 - \hat{z}_i^k} \ y_{5,5} \ v^{100}]_k^+ rem]_0^0 $$

    Objects $err$ are created only in case some overflow appears. Seven transition steps later, the objects $EXIT$ will be in the skin. These objects will represent the final result of $z(t)$ for each $t$: 
    \begin{equation*}
     \hspace*{-0.7cm}   \mathbb{C}_{\leq 130} = [ [ [ \langle AUX, 1 \rangle  \ ]^0_{RES_{i,k}}  \ \langle INIT, k, i, l \rangle^{\hat{z}_i^k}  ]^0_{S_{i,k}} ]_k^0 \ \langle EXIT, k, i, l, 1 \rangle^{\hat{z}_i^k} \ y_{5,12,1} \dots y_{5,12,N} ]_0^0
    \end{equation*}
    Finally, after six more iterations:
    $$ \mathbb{C}_{\leq 136} = [ [ [ \langle AUX, 1 \rangle  \ ]^0_{RES_{i,k}}  \ \langle k, i, l \rangle^{\hat{z}_i^k} ]^0_{S_{i,k}} rem ]_k^0 \ \langle EXIT, k, i, l, 1 \rangle^{\hat{z}_i^k}   [  y_0 ]_P^0]_0^0 $$

In conclusion, the transition from $z(t)$ to $z(t+0.01)$ has a constant upper bound of 136 steps, so the complexity of the system is linear with respect to the time, and it does not depend on the size of the problem. It is important to notice that the last configuration of the system allows the computation of another whole loop of Algorithm \ref{alg:psystem}. 

For this system, we have chosen a particular $f(z)$ to compute $p(z)$ and to use objects that represent $1\%$ of the objects. Because of that, we round everything to two decimals. If we take smaller steps, then it will probably require more time steps to find the GNE, but as said before, the system will only grow linearly with respect to time.

\subsection{Experiment}

Researchers have developed different P system simulators that are available, like P-lingua (MeCoSim) \cite{DBLP:journals/nc/Gutierrez-NaranjoPR08, MecosimRef16, MecosimRef17} or UPSimulator \cite{UPSimulator, UPSimulator2}. To perform experiments, UPSimulator has been chosen because of its flexibility. This simulator allows the design of P systems with particular properties, in this case, transition P systems with membrane polarization. 

\begin{figure}
    \centering
    \includegraphics[width=0.75\linewidth]{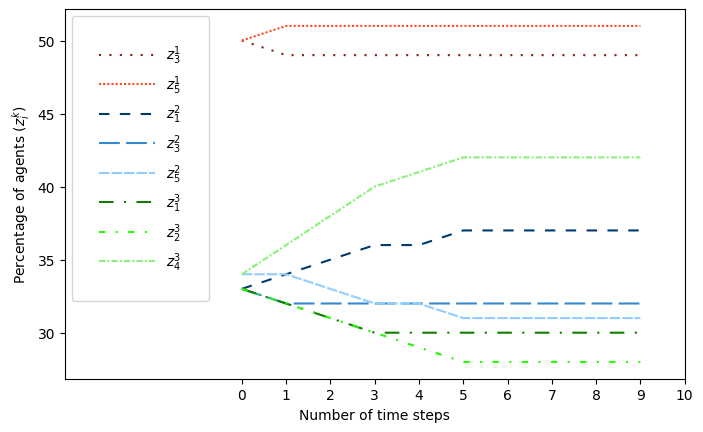}
    \caption{Small experiment with three players and five strategies, distributed such that $S^1 = \{3,5\}$, $S^2 = \{1,3,5\}$, $S^3 = (S^1)^c = \{1,2,4\}$. Players 1, 2, and 3 are represented by red, blue, and green lines respectively. After five time steps, the GNE is reached. It was expected that player 3 would concentrate more agents over time slot 4 ($z_4^3$) or 2 ($z_2^3$) since it is the only player with access to those resources. The difference is a consequence of the cost difference. The key observation in this plot is that $\dot{z} = 0$ after five time steps.}
    \label{fig:GNEexperimento}
\end{figure}

To test the correct behavior of our P system, a small example has been tested, where $N=3$, $T=5$, $S^1 = \{3,5\}$, $S^2 = \{1,3,5\}$, $S^3 = (S^1)^c = \{1,2,4\}$, and we randomly sample the nonzero elements of $D$, $\bar{J}$, $\alpha$, $\beta$ and $m^k$ from $[0,1]$, $[2,4]$, $[1,10]$, $[0,1]$ and $[3,4]$ respectively. In Fig. \ref{fig:GNEexperimento} we can see how the trajectories of the values $z_i^k$ initialize (almost) equally distributed for each player $k$, and they evolve until reaching a stable distribution after only 5 time steps. At that moment, a GNE is found.

\section{Conclusions}\label{section:conclusions}

A P system that computes approximations of GNE for population games under BNN dynamics is proposed. We have proved that the complexity of the system is linear with respect to time, independently of the size of the problem. 

Some of the evolution rules of the P system are defined specifically for instances of the Energy Market Game, as representant of population games under BNN dynamics. However, the P system can be easily modified to compute GNE for other population games under BNN dynamics by changing the payoff function and modifying the corresponding rules. This P system is also easily generalizable to study the evolution of other EDM-PDM systems, even if equilibria are not reached. Future work should focus on expanding the exchange between Game Theory and Membrane Computing.

\section*{Acknowledgments}
M.A.G.N. acknowledges the supported by the European Union HORIZON-CL4-2021-HUMAN-01-01 under grant agreement 101070028 (REXASI-PRO) and by  
TED2021-129438B-I00 / AEI/10.13039/501100011033 / Unión Europea NextGenerationEU/PRTR.

\appendix
\section{Russian peasant multiplication in Membrane Computing \label{appendix:russian}}
A P system that computes the product of two numbers $m,n \in \mathbb{N}_0$, inspired by the Russian peasant multiplication algorithm \cite{Cameron-1994}, is defined in this appendix. The structure of this P system only has three membranes: $\mu = [[]_1[]_2]_0$, and the initial multisets are $m_1 = \{a^m k_1\}$, $m_2 = \emptyset$ and $m_0 = \{b^n\}$.

The ruleset of this system is the following: \\
$RS_1 = [k_1 \rightarrow k_2 \ y_0]_1^0$ \\
$RS_2 = [k_2 \rightarrow k_3]_1^0$ \\
$RS_3 = [k_3 \rightarrow k_4]_1^0$ \\
$RS_4 = [k_4 \rightarrow k_5]_1^0$ \\
$RS_5 = [k_5 \rightarrow k_6]_1^0$ \\
$RS_6 = [k_6 \rightarrow k_1]_1^0$ \\
$RS_7 = [a^2 \rightarrow a_1 \ y_1^2]_1^0$ \\
$RS_8 = [a \rightarrow m \ y_1]_1^0$,  with $\rho_7 > \rho_8$ \\
$RS_9 = [a_1 \rightarrow a_2]_1^0$ \\
$RS_{10} = [a_2 \rightarrow a_3]_1^0$ \\
$RS_{11} = [a_3 \rightarrow a_4]_1^0$ \\
$RS_{12} = [a_4 \rightarrow a_5]_1^0$ \\
$RS_{13} = [a_5 \rightarrow a]_1^0$ \\
$RS_{14} = [y_1^2 \ y_0 \rightarrow \lambda]_1^0$ \\
$RS_{15} = [y_1 \ y_0 \ m]_1^0 \rightarrow [ f ]_1^0 f$, with $\rho_{14} > \rho_{15}$ \\
$RS_{16} = [y_1 \rightarrow \lambda]_1^0$, with $\rho_{15} > \rho_{16}$ \\
$RS_{17} = [m]_1^0 \rightarrow [ \ ]_1^0 m$, with $\rho_{15} > \rho_{17}$ \\
$RS_{18} = [k_2 y_0]_1^0 \rightarrow [ \ ]_1^0 y_0$, with $\rho_{15} > \rho_{18}$ and $\rho_{18} > \rho_{2}$\\
$RS_{19} = [b \rightarrow b_1]_0^0$ \\
$RS_{20} = [b_1 \rightarrow b_2]_0^0$ \\
$RS_{21} = [b_2 \rightarrow b_3]_0^0$ \\
$RS_{22} = [b_3 \rightarrow b_4]_0^0$ \\
$RS_{23} = [b_4 \rightarrow c^2]_0^0$ \\
$RS_{24} = [c \rightarrow b]_0^0$ \\
$RS_{25} = [b_3 [ \ ]_2^0 \rightarrow [d]_2^0 c^2]_0^+$ \\
$RS_{26} = [c \rightarrow b]_0^+$ \\
$RS_{27} = [m]_0^0 \rightarrow [m_1]_0^+ rem$ \\ 
$RS_{28} = [m_1 \rightarrow m_2]_0^+$ \\
$RS_{29} = [m_2 \rightarrow m_3]_0^+$ \\
$RS_{30} = [m_3]_0^+ \rightarrow [ \ ]_0^0 rem$ \\
$RS_{31} = [f \ k_3 \rightarrow \lambda ]_1^0$, with $\rho_{31} > \rho_{3}$ \\
$RS_{32} = f [ \ ]_2^0  \rightarrow f_1 \ [f_1 ]_2^0$ \\
$RS_{33} = [f_1]_2^0 \rightarrow [ \ ]_2^- rem$ \\
$RS_{34} =  [ f_1 ]_0^0  \rightarrow [f_2 ]_0^- rem$ \\ 
$RS_{35} = f_2 [ \ ]_2^-  \rightarrow f_3 \ [f_3 ]_2^-$ \\
$RS_{36} = [f_3 \rightarrow f_4]_2^-$ \\
$RS_{37} = [f_3 \rightarrow f_4]_0^-$ \\
$RS_{38} = [f_4]_2^- \rightarrow [ \ ]_2^0 rem$ \\
$RS_{39} = [f_4]_0^- \rightarrow [ \ ]_0^0 f$ \\
$RS_{40} = [d]_2^- \rightarrow [ \ ]_2^- d$ \\
$RS_{41} = [d]_0^- \rightarrow [ \ ]_0^- d$ \\
$RS_{42} = [b_4 \rightarrow d]_0^-$ \\
$RS_{43} = [y_0]_0^0 \rightarrow [ f_3 ]_0^- rem$ \\ 
$RS_{44} = [b_3 \rightarrow \lambda]_0^-$ \\

\subsection{Computation Analysis}
Four cases will be considered: $m=0$, $m=1$, $m = 2p$ with $p \geq 1$ and $m = 2p + 1$ with $p \geq 1$.
\begin{itemize}
    \item If $m = 0$, then $m \cdot n = 0$, and the following computation takes place:
    \begin{align*}
        \mathbb{C}_0 = [[k_1]_1^0 [ \ ]_2^0 b^n]_0^0, && \\
        \mathbb{C}_1 = [[k_2 y_0]_1^0 [ \ ]_2^0 b_1^n]_0^0, && \  (RS_{1}, RS_{19}) \\
        \mathbb{C}_2 = [[ \ ]_1^0 [ \ ]_2^0 b_2^n y_0]_0^0, && \  (RS_{18}, RS_{20}) \\
        \mathbb{C}_3 = [[ \ ]_1^0 [ \ ]_2^0 b_3^n f_3]_0^- rem, && \ (RS_{21}, RS_{43}) \\
        \mathbb{C}_4 = [[ \ ]_1^0 [ \ ]_2^0 f_4]_0^-, && \  (RS_{36}, RS_{44}) \\
        \mathbb{C}_5 = [[ \ ]_1^0 [ \ ]_2^0 ]_0^0 f, && \  (RS_{39}) 
    \end{align*}
    The computation is then done in 5 iterations, and because the output is zero copies of object $d$, the result is correct.

    \item If $m = 1$, then $m\cdot n = n$ and: 
    \begin{align*}
        \mathbb{C}_0 = [[a k_1]_1^0 [ \ ]_2^0 b^n]_0^0, && \\
        \mathbb{C}_1 = [[m y_1 y_0 k_2]_1^0 [ \ ]_2^0 b_1^n]_0^0, && \  (RS_{1}, RS_{8}, RS_{19}) \\
        \mathbb{C}_2 = [[ f k_3 ]_1^0 [ \ ]_2^0 b_2^n f]_0^0, && \  (RS_{2}, RS_{15}, RS_{20}) \\
        \mathbb{C}_3 = [[ \ ]_1^0 [ f_1 ]_2^0 b_3^n f_1]_0^0, && \ (RS_{21}, RS_{31}, RS_{32}) \\
        \mathbb{C}_4 = [[ \ ]_1^0 [ \ ]_2^- b_4^n f_2 rem]_0^- rem, && \ (RS_{22}, RS_{33}, RS_{34}) \\
        \mathbb{C}_5 = [[ \ ]_1^0 [ f_3 ]_2^- d^n f_3 ]_0^- , && \ (RS_{35}, RS_{42}) \\
        \mathbb{C}_6 = [[ \ ]_1^0 [ f_4 ]_2^- f_4 ]_0^- d^n  , && \ (RS_{36}, RS_{37}, RS_{41}) \\
        \mathbb{C}_7 = [[ \ ]_1^0 [ \ ]_2^0 rem ]_0^0 f d^n  , && \ (RS_{38}, RS_{39}) \\
    \end{align*}
    The computation is then done in 7 iterations, and because the output is $n$ copies of object $d$, the result is correct.

    \item If $m = 2p$ with $p\geq 1$ and we follow the Russian peasant multiplication algorithm, then the result of the partial computation doesn't contribute directly to the final result. In our system, this means that no objects $d$ will be saved in membrane $[ \ ]_2$:
    \begin{align*}
        \mathbb{C}_0 = [[a^{2p} k_1]_1^0 [ \ ]_2^0 b^n]_0^0, && \\
        \mathbb{C}_1 = [[a_1^p y_1^{2p} y_0 k_2]_1^0 [ \ ]_2^0 b_1^n]_0^0, && \  (RS_{1}, RS_{7}, RS_{19}) \\
        \mathbb{C}_2 = [[ a_2^p k_3 ]_1^0 [ \ ]_2^0 b_2^n ]_0^0, && \  (RS_{2}, RS_{9}, RS_{14}, RS_{16}) \\
        \mathbb{C}_3 = [[ a_3^p k_4 ]_1^0 [ \ ]_2^0 b_3^n]_0^0, && \ (RS_{3}, RS_{10}, RS_{21}) \\
        \mathbb{C}_4 = [[ a_4^p k_5 ]_1^0 [ \ ]_2^0 b_4^n ]_0^0, && \ (RS_{4}, RS_{11}, RS_{22}) \\
        \mathbb{C}_5 = [[ a_5^p k_6 ]_1^0 [ \ ]_2^0 c^{2n} ]_0^0, && \ (RS_{5}, RS_{12}, RS_{23}) \\
        \mathbb{C}_6 = [[ a^p k_1 ]_1^0 [ \ ]_2^0 b^{2n} ]_0^0, && \ (RS_{6}, RS_{13}, RS_{24}) \\
    \end{align*}
    The computation goes from multiplying by $m=2p$ to multiplying by $m=p$ in 6 iterations. We can see also that no objects $d$ were created, and that $n$ raised to $2n$ as in the Russian peasant multiplication algorithm.

    \item If $m = 2p+1$ with $p\geq 1$ and we follow the Russian peasant multiplication algorithm, then in this case the result of the partial computation contributes directly to the final result. In our system, this means that $n$ objects $d$ will be saved in membrane $[ \ ]_2$:
    \begin{align*}
        \mathbb{C}_0 = [[a^{2p+1} k_1]_1^0 [ \ ]_2^0 b^n]_0^0, && \\
        \mathbb{C}_1 = [[a_1^p \ m \ y_1^{2p+1} \ y_0 \ k_2]_1^0 [ \ ]_2^0 b_1^n]_0^0, && \  (RS_{1}, RS_{7}, RS_{8}, RS_{19}) \\
        \mathbb{C}_2 = [[ a_2^p k_3 ]_1^0 [ \ ]_2^0 b_2^n \ m]_0^0, && \  (RS_{2}, RS_{9}, RS_{14}, RS_{16}, RS_{17}) \\
        \mathbb{C}_3 = [[ a_3^p k_4 ]_1^0 [ \ ]_2^0 b_3^n \ m_1]_0^+ rem, && \ (RS_{3}, RS_{10}, RS_{21}, RS_{27}) \\
        \mathbb{C}_4 = [[ a_4^p k_5 ]_1^0 [ d^n ]_2^0 c^{2n} \ m_2 ]_0^+, && \ (RS_{4}, RS_{11}, RS_{25}, RS_{28}) \\
        \mathbb{C}_5 = [[ a_5^p k_6 ]_1^0 [ \ ]_2^0 b^{2n} \ m_3 ]_0^+, && \ (RS_{5}, RS_{12}, RS_{26}, RS_{29}) \\
        \mathbb{C}_6 = [[ a^p k_1 ]_1^0 [ \ ]_2^0 b^{2n} ]_0^0 rem, && \ (RS_{6}, RS_{13}, RS_{30}) \\
    \end{align*}
    The computation goes from multiplying by $m=2p+1$ to multiplying by $m=p$ in 6 iterations. We can see also that $n$ objects $d$ were created in membrane $[ \ ]_2^0$, and that $n$ raised to $2n$ as in the Russian peasant multiplication algorithm.
\end{itemize}

Finally, if we observe rules $RS_{40}$ and $RS_{41}$, we see that the objects $d$ will be extracted from $[ \ ]_2^0$ when the whole computation is finished, giving the result as output.

It takes $6$ iterations to reduce the problem of multiplying by $m$ to the problem of multiplying by $\lfloor m/2 \rfloor$. Because $n$ doesn't influence the number of operations, we have an upper bound over the number of iterations of $1 + 6 \cdot \lceil log_2(m) \rceil$.

We use this fact by placing $z_i^k$ as $m$, and because $1 \leq z_i^k \leq 100$, we have that if we multiply $m$ by any number, the number of iterations will be bounded by $1 + 6 \cdot \lceil log_2(100) \rceil = 43$.

In this article, the membranes $MULT_{i,k}$ will be almost exactly as described, with the structure changing from $\mu = [[ \ ]_1^0 [ \ ]_2^0]_0^0$ to $\mu' = [[ \ ]_{M1}^0 [ \ ]_{M2}^0]_{MULT_{i,k}}^0$. Membranes $MULT2_{i,k}$ will have $\mu'' = [[ \ ]_{M1'}^0 [ \ ]_{M2'}^0]_{MULT2_{i,k}}^0$ as membrane structure, and will return objects $d_i$ instead of $d$ and objects $f_1$ instead of $f$.

\section{Definition of the P system \label{appendix:DefinitionPSystem}}

Let us consider the following transition P system with membrane polarization $$ \Pi = \langle \Gamma, H, EC, \mu, \{w_h\}_{h \in H}, (\mathcal{R},\rho) \rangle $$
where the alphabet of objects is given by:
\begin{align*}
    \Gamma &= \{ \langle Prod, k,i,l \rangle, \langle Prod2, k,i,l \rangle, \langle k,i,l \rangle, C_k, \langle a,k,i,l \rangle \ | \ k \in \mathcal{P}, i \in S^k, l = \sum\limits_{j < k} |S^j| + i \}   \\
    &\cup \{ c, rem, mult_0, prod, e, pos, q, s_0, s_1, zneg, zvarp, zvarn, a, b, d, m, f, y_0, p, n, comp \} \\
    &\cup \{ m_i, k_i, a_i, b_i, f_i, y_i \ | \ 1 \leq i \leq 6 \} \cup \{ over, p_1, err, v\}\\  
    & \cup \{y_{7,k} \ | \ k \in \mathcal{P}\} \cup \{ y_{0,0}, y_{0,1}, y_{0,2}, y_{2,0}, y_{2,1}, y_{2,2}, y_{3,0}, y_{4,0}, y_{4,1}, y_{4,2}, y_{4,3}, y_{5,0} \} \\
    & \cup \{ p_l \ | \ 1 \leq l \leq  \sum\limits_{k \in \mathcal{P}} |S^k|\} \cup \{ \langle AUX, n \rangle, \langle AUX1, n \rangle, \langle CLK, n \rangle | n \geq 0 \}\\
    & \cup \{ y_{3,j,i}, multz_{i,0}, multz_{i,1}, \langle q, i \rangle, d_i, q_i, neg_i, pos_i \ | \ k \in \mathcal{P}, i \in S^k , 1 \leq j \leq 7\} \\
    & \cup \{ y_{0,k}, y_{0,i}, y_{5,j}, y_{5,3,i}, y_{5,11,i}, y_{5,12,k}, w_i, compw_i, z_i, \ | \ k \in \mathcal{P}, i \in S^k , 1 \leq j \leq 10 \} \\
    & \cup \{ EXIT_i, \langle EXIT,k,i,l,n \rangle, \langle INIT,k,i,l \rangle\ | \ k \in \mathcal{P}, i \in S^k , 1 \leq j \leq 10, n \geq 1 \}
\end{align*}

The set of membrane labels is given by $H = \{0\} \ \cup \ \mathcal{P} \ \cup \ \{S_{i,k}\}_{\forall i \in S^k \forall k \in \mathcal{P}} \ \cup \ \{RES_{i,k}\}_{\forall i \in S^k \forall k \in \mathcal{P}} \ \cup \ \{MULT_{i,k}\}_{\forall i \in S^k \forall k \in \mathcal{P}} \ \cup \ \{M1, M2\} \ \cup \ \{MULT2_{i,k}\}_{\forall i \in S^k \forall k \in \mathcal{P}} \ \cup \ \{UPD_{i,k}\}_{\forall i \in S^k \forall k \in \mathcal{P}} \ \cup \ \{ACUM_{k}\}_{\forall k \in \mathcal{P}}$

The set of electrical charges is given by $EC = \{0, +, -\}$.

The membrane structure $\mu$ can be defined as follows: 
    \begin{itemize}
        \item Membrane skin with label $0$, inside of which:
        \begin{enumerate}[label*=\arabic*.]
            \item One membrane with label $P$.
            \item $N$ membranes with labels $\mathcal{P}$. Inside of each membrane $k \in \mathcal{P}$:
            \begin{enumerate}[label*=\arabic*.]
                \item $S^k$ membranes with labels $S_{i,k}$ $\forall i \in S^k$. Inside each $S_{i,k}$:
                \begin{itemize}
                    \item One membrane with label $RES_{i,k}$
                \end{itemize}
                \item $S^k$ membranes with labels $MULT_{i,k}$ $\forall i \in S^k$. Inside each membrane $MULT_{i,k}$:
                \begin{itemize}
                    \item One membrane with label $M1$
                    \item One membrane with label $M2$
                \end{itemize}
                \item $S^k$ membranes with labels $MULT2_{i,k}$ $\forall i \in S^k$. Inside each membrane $MULT2_{i,k}$:
                \begin{itemize}
                    \item One membrane with label $M1'$
                    \item One membrane with label $M2'$
                \end{itemize}
                \item $S^k$ membranes with labels $UPD_{i,k}$ $\forall i \in S^k$
                \item One membrane with label $ACUM_{k}$
            \end{enumerate}
            
        \end{enumerate}
    \end{itemize}

The initial multisets are $w_P = {y_0}$, $w_{S_{i,k}} = \langle k,i,l \rangle^{z_i^k}$ $\forall i \in S^k$ $\forall k \in \mathcal{P}$ with $z_i^k := \lfloor \frac{100}{n^k} \rfloor$ for $1 \leq i < \max S^k$ and $z_{max S^k}^k := 100 - (|S^k| - 1) \cdot \lfloor \frac{100}{n^k} \rfloor$, $w_{RES_{i,k}} = \langle AUX, 0 \rangle$ $\forall i \in S^k$ $\forall k \in \mathcal{P}$, and for any other membrane $m$, the initial multiset is $w_m = \emptyset$

The set of rules $\mathcal{R}$ is given by the following rules, separated by stage, where $\lambda$ represents the empty multiset, the rule $RS_{m,n}$ represents the $n$-th rule of the $m$-th stage, the rules are defined $\forall k \in \mathcal{P}$, $\forall i \in S^k$, and $l = \sum\limits_{j < k} |S^j| + i$, and $\rho_{m,n}$ represents the priority of the rule $RS_{m,n}$: \\

\textbf{Stage 1 (Computation of payoff $p(t)$)} To use $f(z(t))$ in the P system, let $\kappa := \lfloor 100 \cdot (- C^\top \bar{J} - \beta) \rfloor$ be the constant part, and let $a_{j,l} := \lfloor (S \cdot M)_{jl} \rfloor$ and $b_{l} := \lfloor (S \cdot M)_{ll} + (\alpha \cdot M)_{l} \rfloor$ for $1 \leq j,l \leq \sum\limits_{k \in \mathcal{P}} |S^k|$ be the variable part. Because  $a_{j,l}$ and $b_l$ represent the contribution of $1\%$ of the agents, the rounding coefficient $100$ is canceled. Let also $n = \sum\limits_{k \in \mathcal{P}} |S^k|$: \\
        $RS_{1,1} \equiv [\langle k,i,l \rangle]^0_{S_{i,k}} \rightarrow [c]^0_{S_{i,k}} \langle k,i,l \rangle$ \\
        $RS_{1,2} \equiv [y_0 \rightarrow  p_1^{\kappa_1} p_2^{\kappa_2} \dots p_n^{\kappa_n} ]^0_P$  \\
        $RS_{1,3} \equiv [\langle k,i,l \rangle]^0_{k} \rightarrow [\langle Prod,k,i,l \rangle]^0_{k} \langle k,i,l \rangle C_k$ \\
        $RS_{1,4} \equiv \langle k,i,l \rangle [ \ ]^0_P \rightarrow [\langle k,i,l \rangle]]^0_P$\\
        $RS_{1,5} \equiv [C_1^{100} C_2^{100} \dots C_N^{100} \rightarrow  y_1 ]^0_0$ \\ 
        $RS_{1,6} \equiv [y_1 [ \ ]^0_P \rightarrow  [y_2]_P^+ ]^0_0$ \\
        $RS_{1,7} \equiv [\langle k,i,l \rangle \rightarrow p_1^{a_{1,l}} p_2^{a_{2,l}} \dots p_{l-1}^{a_{l-1,l}} p_l^{b_l} p_{l+1}^{a_{l+1,l}} \dots p_n^{a_{n,l}}]^+_P$\\
        $RS_{1,8} \equiv [y_2 \rightarrow y_3]_P^+$ \\
        $RS_{1,9} \equiv [y_3]_P^+ \rightarrow [y_4]_P^- rem$  \\
        $RS_{1,10} \equiv [p_l]_P^- \rightarrow [ \ ]^-_P \langle a,k,i,l \rangle$ \\
        $RS_{1,11} \equiv [y_4 \rightarrow y_5]_P^-$ \\
        $RS_{1,12} \equiv [\langle a,k,i,l \rangle[ \ ]_k^0 \rightarrow [ \langle a,k,i,l \rangle]_k^0 ]^0_0 $ \\
        $RS_{1,13} \equiv [y_5 \rightarrow y_6]_P^-$ \\
        $RS_{1,14} \equiv [y_6]_P^- \rightarrow [ \ ]_P^0 \  y_{7,1}  \ y_{7,2}  \ ... \  y_{7,N}$ \\
        $RS_{1,15} \equiv [y_{7,k}[ \ ]_k^0 \rightarrow [ mult_0^{|S^k|}]_k^- ]^0_0 $, $\forall k \in \mathcal{P}$. \\
        $RS_{1,16} \equiv [rem \rightarrow \lambda]_m^s$, $\forall m \in H$, $\forall s \in \{+,-,0\}$. \\

\textbf{Stage 2 (Computation of sums $\sum\limits_{j \in S^k} z_j^k(t) p_j^k(t)$)} \\
        $RS_{2,1} \equiv [\langle Prod,k,i,l \rangle [ \ ]^0_{MULT_{i,k}} \rightarrow [ prod ]^0_{MULT_{i,k}} \langle Prod2,k,i,l \rangle]^-_k $\\
        $RS_{2,2} \equiv [\langle a,k,i,l \rangle [ \ ]^0_{MULT_{i,k}} \rightarrow [ e ]^0_{MULT_{i,k}} neg_i]^-_k $\\
        $RS_{2,3} \equiv [mult_0 [ \ ]^0_{MULT_{i,k}} \rightarrow [ mult_0 ]^+_{MULT_{i,k}}]^-_k $ \\
        $RS_{2,4} \equiv [prod \rightarrow [ a ]^0_{M1}]^+_{MULT_{i,k}} $ \\
        $RS_{2,5} \equiv [e \rightarrow b]^+_{MULT_{i,k}} $ \\
        $RS_{2,6} \equiv [mult_0 [ \ ]_{M1}^0]^+_{MULT_{i,k}} \rightarrow [[k_1]_{M1}^0]^0_{MULT_{i,k}} rem $ \\
        $RS_{2,7} \equiv [neg_i [ \ ]^0_{UPD_{i,k}} \rightarrow [ neg_i ]^0_{UPD_{i,k}} ]^-_k $ \\
        $RS_{2,8} \equiv [f^{|S^k|} [ \ ]^0_{ACUM_{k}} \rightarrow [ y_{2,0} ]^+_{ACUM_{k}}]^-_k $ \\
        $RS_{2,9} \equiv [d [ \ ]^0_{ACUM_{k}} \rightarrow [ pos ]^0_{ACUM_{k}}]^-_k $ \\
        $RS_{2,10} \equiv [ pos^{100} ]^+_{ACUM_{k}} \rightarrow [ \ ]^+_{ACUM_{k}} pos $ \\
        $RS_{2,11} \equiv [ pos^{51} ]^+_{ACUM_{k}} \rightarrow [ \ ]^+_{ACUM_{k}} pos $, with $\rho_{2,10} > \rho_{2,11}$. \\
        $RS_{2,12} \equiv [ pos \rightarrow \lambda ]^+_{ACUM_{k}} $,  with $\rho_{2,11} > \rho_{2,12}$. \\
        $RS_{2,13} \equiv [ y_{2,0} \rightarrow y_{2,1} ]^+_{ACUM_{k}} $ \\
        $RS_{2,14} \equiv [ y_{2,1} ]^+_{ACUM_{k}} \rightarrow [ \ ]^0_{ACUM_{k}} y_{2,2} $ \\
        $RS_{2,15} \equiv [ y_{2,2} ]^-_{k} \rightarrow [ y_{3,0} ]^0_{k} rem $ \\

\textbf{Stage 3 (Computation of $[ \hat{p}_i^k ]_+$ and $\sum\limits_{j \in S^k} [\hat{p}_j^k ]_+$)}
        
        Let $S^k = \{ i_1, \dots, i_{|S^k|} \} $. Then: \\
        $RS_{3,1} \equiv [y_{3,0} \rightarrow y_{3,1,i_{1}} \ \dots \ y_{3,1,i_{|S^k|}}]^0_k $ \\
        $RS_{3,2} \equiv [y_{3,1,i} [ \ ]^0_{UPD_{i,k}} \rightarrow [ rem ]^+_{UPD_{i,k}} y_{3,2,i} ]^0_k $ \\
        $RS_{3,3} \equiv [pos \rightarrow pos_{i_{1}} \ \dots \ pos_{i_{|S^k|}} ]^0_k $ \\
        $RS_{3,4} \equiv [pos_{i} [ \ ]^+_{UPD_{i,k}} \rightarrow [ pos_{i} ]^+_{UPD_{i,k}} ]^0_k $ \\
        $RS_{3,5} \equiv [y_{3,2,i} \rightarrow y_{3,3,i} ]^0_k $ \\
        $RS_{3,6} \equiv [y_{3,3,i} [ \ ]^+_{UPD_{i,k}} \rightarrow [ y_{3,4,i} ]^-_{UPD_{i,k}} ]^0_k $ \\
        $RS_{3,7} \equiv [neg_i \ pos_i  \rightarrow \lambda ]^-_{UPD_{i,k}}  $ \\
        $RS_{3,8} \equiv [neg_i  \rightarrow \lambda ]^-_{UPD_{i,k}}  $, with $\rho_{3,7} > \rho_{3,8}$. \\
        $RS_{3,9} \equiv [pos_i  \rightarrow q_i ]^-_{UPD_{i,k}}  $, with $\rho_{3,7} > \rho_{3,9}$. \\
        $RS_{3,10} \equiv [y_{3,4,i} \rightarrow y_{3,5,i} ]^-_{UPD_{i,k}} $ \\
        $RS_{3,11} \equiv [y_{3,5,i}]^-_{UPD_{i,k}} \rightarrow [y_{3,6,i} ]^0_{UPD_{i,k}} rem $ \\
        $RS_{3,12} \equiv [q_i]^0_{UPD_{i,k}} \rightarrow [ \ ]^0_{UPD_{i,k}} q \ q_i $ \\
        $RS_{3,13} \equiv [y_{3,6,i}]^0_{UPD_{i,k}} \rightarrow [ \ ]^0_{UPD_{i,k}} y_{3,7,i} $ \\

\textbf{Stage 4 (Computation of $\dot{z}_i^k(t)$)}  Let $S^k = \{ i_1, \dots, i_{|S^k|} \} $. Then: \\
        $RS_{4,1} \equiv [y_{3,7,i_1} \dots y_{3,7,i_{|S^k|}} \rightarrow multz_{i_1, 0} \dots multz_{i_{|S^k|}, 0}]^0_k $ \\
        $RS_{4,2} \equiv [multz_{i, 0} \rightarrow multz_{i, 1}]^0_k $ \\
        $RS_{4,3} \equiv [q \rightarrow \langle q,i_{1} \rangle \dots \langle q,i_{|S^k|} \rangle]^0_k $ \\
        $RS_{4,4} \equiv [\langle Prod2,k,i,l \rangle [ \ ]^0_{MULT2_{i,k}} \rightarrow [ prod ]^0_{MULT2_{i,k}} ]^0_k $ \\
        $RS_{4,5} \equiv [\langle q,i \rangle [ \ ]^0_{MULT2_{i,k}} \rightarrow [ e ]^0_{MULT2_{i,k}} ]^0_k $ \\
        $RS_{4,6} \equiv [multz_{i, 1} [ \ ]^0_{MULT2_{i,k}} \rightarrow [ mult_0 ]^+_{MULT2_{i,k}} ]^0_k $ \\
        $RS_{4,7} \equiv [prod \rightarrow [ a ]^0_{M1'}]^+_{MULT2_{i,k}} $ \\
        $RS_{4,8} \equiv [e \rightarrow b]^+_{MULT2_{i,k}} $ \\
        $RS_{4,9} \equiv [mult_0 [ \ ]_{M1'}^0]^+_{MULT2_{i,k}} \rightarrow [[k_1]_{M1'}^0]^0_{MULT2_{i,k}} rem $ \\
        $RS_{4,10} \equiv [f_1^{|S^k|} \rightarrow y_{4,0}^{|S^k|} ]^0_k $ \\
        $RS_{4,11} \equiv [y_{4,0} [ \ ]^0_{S_{i,k}} \rightarrow [ y_{4,1} ]^+_{S_{i,k}} ]^0_{k} $ \\ 
        $RS_{4,12} \equiv [q_i [ \ ]^+_{S_{i,k}} \rightarrow [ s_0 ]^+_{S_{i,k}} ]^0_{k} $ \\
        $RS_{4,13} \equiv [ d_i [ \ ]^+_{S_{i,k}} \rightarrow [ d_i ]^+_{S_{i,k}}]_k^0 $ \\
        $RS_{4,14} \equiv [ s_0 \rightarrow s_1 ]^+_{S_{i,k}}  $ \\
        $RS_{4,15} \equiv [ d_i^{100} \rightarrow zneg ]^+_{S_{i,k}}  $ \\
        $RS_{4,16} \equiv [ d_i^{51} \rightarrow zneg ]^+_{S_{i,k}} $, with $\rho_{4,15} > \rho_{4,16}$ . \\
        $RS_{4,17} \equiv [ d_i \rightarrow \lambda ]^+_{S_{i,k}} $, with $\rho_{4,16} > \rho_{4,17}$. \\
        $RS_{4,18} \equiv [ s_1 zneg \rightarrow \lambda ]^+_{S_{i,k}} $ \\
        $RS_{4,19} \equiv [ s_1 \rightarrow zvarp ]^+_{S_{i,k}} $, with $\rho_{4,18} > \rho_{4,19}$. \\
        $RS_{4,20} \equiv [ zneg \rightarrow zvarn ]^+_{S_{i,k}} $, with $\rho_{4,18} > \rho_{4,20}$. \\
        $RS_{4,21} \equiv [ y_{4,1} \rightarrow y_{4,2} ]^+_{S_{i,k}} $ \\
        $RS_{4,22} \equiv [ y_{4,2} \rightarrow y_{4,3} ]^+_{S_{i,k}} $  \\
        $RS_{4,23} \equiv [ y_{4,3}]^+_{S_{i,k}} \rightarrow [y_{5,0} ]^-_{S_{i,k}} rem $ \\

\textbf{Stage 5 (Update of $z(t)$, coordination for next iteration and results output)} \\
        $RS_{5,1} \equiv [y_{5,0} \rightarrow y_{5,1}]^-_{S_{i,k}} $ \\
        $RS_{5,2} \equiv [zvarn^{100} \ c \rightarrow \lambda]^-_{S_{i,k}} $ \\
        $RS_{5,3} \equiv [zvarn^{51} \ c \rightarrow \lambda]^-_{S_{i,k}} $, with $\rho_{5,2} > \rho_{5,3}$. \\
        $RS_{5,4} \equiv [zvarp^{100} \rightarrow p]^-_{S_{i,k}} $ \\
        $RS_{5,5} \equiv [c \rightarrow p]^-_{S_{i,k}} $, with $\rho_{5,3} > \rho_{5,5}$.  \\
        $RS_{5,6} \equiv [zvarn^{100} \rightarrow n]^-_{S_{i,k}} $, with $\rho_{5,3} > \rho_{5,6}$. \\
        $RS_{5,7} \equiv [zvarn^{51} \rightarrow n]^-_{S_{i,k}} $, with $\rho_{5,6} > \rho_{5,7}$. \\
        $RS_{5,8} \equiv [zvarn \rightarrow \lambda]^-_{S_{i,k}} $, with $\rho_{5,7} > \rho_{5,8}$. \\
        $RS_{5,9} \equiv [zvarp^{51} \rightarrow p]^-_{S_{i,k}} $, with $\rho_{5,4} > \rho_{5,9}$. \\
        $RS_{5,10} \equiv [zvarp \rightarrow \lambda]^-_{S_{i,k}} $, with $\rho_{5,9} > \rho_{5,10}$. \\
        $RS_{5,11} \equiv [y_{5,1} \rightarrow y_{5,2} \ comp^{100}]^-_{S_{i,k}} $ \\
        $RS_{5,12} \equiv [p^{100}]^-_{S_{i,k}} \rightarrow [over]^-_{S_{i,k}} w_i^{100} $ \\
        $RS_{5,13} \equiv [y_{5,2} ]^-_{S_{i,k}} \rightarrow [y_{5,3} ]^0_{S_{i,k}} y_{5,3, i}$ \\
        $RS_{5,14} \equiv [p]^0_{S_{i,k}} \rightarrow [ \ ]^0_{S_{i,k}} p $ \\
        $RS_{5,15} \equiv [over \ comp^{100} \rightarrow \lambda \ ]^-_{S_{i,k}} $ \\
        $RS_{5,16} \equiv [n]^0_{S_{i,k}} \rightarrow [ \ ]^0_{S_{i,k}} n $ \\
        $RS_{5,17} \equiv [comp]^0_{S_{i,k}} \rightarrow [ \ ]^0_{S_{i,k}} compw_i $ \\
        $RS_{5,18} \equiv [p \ comp \rightarrow p_1 ]^-_{S_{i,k}} $, with $\rho_{15} > \rho_{18}$. \\
        $RS_{5,19} \equiv [p_1]^0_{S_{i,k}} \rightarrow [ \ ]^0_{S_{i,k}} w_i $ \\
        $RS_{5,20} \equiv [p \ n \rightarrow \lambda ]^0_k $ \\
        $RS_{5,21} \equiv [p \ compw_i \rightarrow w_i ]^0_k $, with $\rho_{20} > \rho_{21}$. \\
        $RS_{5,22} \equiv [n \ w_i \rightarrow compw_i ]^0_k $, with $\rho_{20} > \rho_{22}$. \\
        $RS_{5,23} \equiv [p \rightarrow err ]^0_k $, with $\rho_{21} > \rho_{23}$. \\
        $RS_{5,24} \equiv [n \rightarrow err ]^0_k $, with $\rho_{22} > \rho_{24}$. \\
        $RS_{5,25} \equiv [y_{5,3} \rightarrow y_{5,4} ]^0_{S_{i,k}} $ \\
        $RS_{5,26} \equiv [y_{5,4} \rightarrow y_{5,5} ]^0_{S_{i,k}} $ \\
        $RS_{5,27} \equiv [y_{5,5} ]^0_{S_{i,k}} \rightarrow [ \ ]^+_{S_{i,k}} y_{5,6}$ \\
        $RS_{5,28} \equiv [y_{5,3,i_1} \dots y_{5,3,i_{|S^k|}} \rightarrow y_{5,4} ]^0_k $ \\
        $RS_{5,29} \equiv [y_{5,4} ]^0_k \rightarrow [ y_{5,5} \ v^{100} ]^+_k rem$ \\
        $RS_{5,30} \equiv [w_i \ v \rightarrow z_i]^+_k $ \\
        $RS_{5,31} \equiv [w_i \rightarrow \lambda]^+_k $, with $\rho_{5,30} > \rho_{5,31}$. \\
        $RS_{5,32} \equiv [compw_i  \rightarrow \lambda ]^+_k $ \\
        $RS_{5,33} \equiv [v  \rightarrow z_i ]^+_k $, with $\rho_{5,30} > \rho_{5,33}$. \\
        $RS_{5,34} \equiv [y_{5,5} ]^+_k \rightarrow [ \ ]^0_k rem$ \\
        $RS_{5,35} \equiv z_i [ \ ]^+_{S_{i,k}} \rightarrow [ z_i ]^+_{S_{i,k}} $ \\
        $RS_{5,36} \equiv y_{5,6} [ \ ]^+_{S_{i,k}} \rightarrow [ y_{5,7} ]^0_{S_{i,k}} $ \\
        $RS_{5,37} \equiv y_{5,7} [ \ ]^0_{RES_{i,k}} \rightarrow [ y_{5,8} ]^+_{RES_{i,k}} $ \\
        $RS_{5,38} \equiv z_i [ \ ]^+_{RES_{i,k}} \rightarrow [ EXIT_i ]^+_{RES_{i,k}} $ \\
        $RS_{5,39} \equiv [ \langle AUX,n\rangle  \rightarrow \langle CLK, n+1\rangle^{100} \langle AUX1,n+1\rangle ]^+_{RES_{i,k}} $ \\
        $RS_{5,40} \equiv [ y_{5,8} ]^+_{RES_{i,k}} \rightarrow [ y_{5,9} ]^-_{RES_{i,k}} rem$ \\
        $RS_{5,41} \equiv [ EXIT_i \ \langle CLK, n \rangle ]^-_{RES_{i,k}} \rightarrow [ \ ]^-_{RES_{i,k}} \langle EXIT, k, i, l, n \rangle $, with $n \geq 1$. \\
        $RS_{5,42} \equiv [ \langle CLK, n \rangle  \rightarrow \lambda ]^-_{RES_{i,k}} $, with $n \geq 1$ and $\rho_{5,41} > \rho_{5,42}$. \\
        $RS_{5,43} \equiv [ y_{5,9} ]^-_{RES_{i,k}} \rightarrow [ \ ]^0_{RES_{i,k}}  y_{5,10}  $ \\
        $RS_{5,44} \equiv [ \langle AUX1, n \rangle  \rightarrow \langle AUX, n \rangle ]^0_{RES_{i,k}} $, with $n \geq 1$ \\
        $RS_{5,45} \equiv [ \langle EXIT, k, i, l, n \rangle ]^0_{S_{i,k}} \rightarrow [ \langle INIT, k, i, l \rangle ]^0_{S_{i,k}}  \langle EXIT, k, i, l, n \rangle $, with $n \geq 1$. \\
        $RS_{5,46} \equiv [ y_{5,10} ]^0_{S_{i,k}} \rightarrow [ \ ]^0_{S_{i,k}}  y_{5,11,i}  $ \\
        $RS_{5,47} \equiv [ \langle EXIT, k, i, l, n \rangle ]^0_k \rightarrow [ \ ]^0_k  \langle EXIT, k, i, l, n \rangle  $, with $n \geq 1$. \\
        $RS_{5,48} \equiv [ y_{5,11,i_1} \dots y_{5,11,i_{|S^k|}} ]^0_k \rightarrow [ \ ]^0_k \ y_{5,12,k}  $\\
        $RS_{5,49} \equiv [ y_{5,12,1} \dots y_{5,12,N}  \rightarrow  y_0 ]^0_0   $\\
        $RS_{5,50} \equiv [ err ]^0_k \rightarrow [ \ ]^0_k  err  $ \\
        $RS_{5,51} \equiv [ y_0 \rightarrow y_{0,0} \ y_{0,1} \dots y_{0,N} ]^0_0  $ \\ 
        $RS_{5,52} \equiv  y_{0,0} [ \ ]_P^0 \rightarrow [y_{0,0}]^0_P  $ \\
        $RS_{5,53} \equiv  y_{0,k} [ \ ]_k^0 \rightarrow [y_{0,0}]^0_k  $ \\
        $RS_{5,54} \equiv  [ y_{0,0}   \rightarrow y_{0,1}]^0_P  $ \\
        $RS_{5,55} \equiv   [ y_{0,0} ]_k^0 \rightarrow [y_{0,i_1} \dots y_{0,i_{|S^k|}}]^0_k  $ \\
        $RS_{5,56} \equiv  [ y_{0,1} \rightarrow y_{0,2}]^0_P  $ \\
        $RS_{5,57} \equiv  y_{0,i} [ \ ]_{S_{i,k}}^0 \rightarrow [y_{0,2}]^+_{S_{i,k}}  $ \\
        $RS_{5,58} \equiv  [ y_{0,2}  \rightarrow y_{0}]^0_P  $ \\
        $RS_{5,59} \equiv  [ \langle INIT, k, i, l \rangle \rightarrow  \langle k, i, l \rangle]^+_{S_{i,k}}  $ \\
        $RS_{5,60} \equiv   [ y_{0,2} ]_{S_{i,k}}^+ \rightarrow [ \ ]^0_{S_{i,k}} rem $ \\
        To stop after $L$ iterations of the loop (Algorithm \ref{alg:psystem}): \\
        $RS_{5,61} \equiv   [ \langle AUX1, L \rangle \ y_{5,9} \rightarrow \lambda ]_{RES_{i,k}}^- $ \\


\begin{thebibliography}{10}
\expandafter\ifx\csname url\endcsname\relax
  \def\url#1{\texttt{#1}}\fi
\expandafter\ifx\csname urlprefix\endcsname\relax\def\urlprefix{URL }\fi
\expandafter\ifx\csname href\endcsname\relax
  \def\href#1#2{#2} \def\path#1{#1}\fi

\bibitem{EGT}
J.~Hofbauer, K.~Sigmund, Evolutionary games and population dynamics, Journal of the American Statistical Association 95 (06 2000).
\newblock \href {https://doi.org/10.2307/2669431} {\path{doi:10.2307/2669431}}.

\bibitem{Nash1951}
J.~Nash, Non-cooperative games, Annals of Mathematics 54~(2) (1951) 286--295.

\bibitem{Debreu1952}
G.~Debreu, A social equilibrium existence theorem, Proceedings of the National Academy of Sciences of the United States of America 38~(10) (1952) 886--893.

\bibitem{DBLP:journals/4or/FacchineiK07}
F.~Facchinei, C.~Kanzow, Generalized {N}ash equilibrium problems, 4OR 5~(3) (2007) 173--210.
\newblock \href {https://doi.org/10.1007/S10288-007-0054-4} {\path{doi:10.1007/S10288-007-0054-4}}.

\bibitem{paunbook}
\relax{Gh}eorghe P\u{a}un, Membrane Computing. {A}n Introduction, Springer-Verlag, Berlin, Germany, 2002.

\bibitem{HandbookMC10}
{\relax Gh}.~P\u{a}un, G.~Rozenberg, A.~Salomaa (Eds.), The {O}xford {H}andbook of {M}embrane {C}omputing, Oxford University Press, Oxford, England, 2010.

\bibitem{scavengers}
M.~Àngels Colomer, A.~Margalida, D.~Sanuy, M.~J. Pérez-Jiménez, A bio-inspired computing model as a new tool for modeling ecosystems: The avian scavengers as a case study, Ecological Modelling 222~(1) (2011) 33--47.
\newblock \href {https://doi.org/https://doi.org/10.1016/j.ecolmodel.2010.09.012} {\path{doi:https://doi.org/10.1016/j.ecolmodel.2010.09.012}}.

\bibitem{chamois}
M.~Colomer, S.~Lavin, I.~Marco, A.~Margalida, I.~Pérez-Hurtado, M.~Pérez-Jiménez, D.~Sanuy, E.~Serrano, L.~Valencia-Cabrera, Modeling population growth of pyrenean chamois (rupicapra p. pyrenaica) by using p-systems, Vol. 6501, 2010, pp. 144--159.
\newblock \href {https://doi.org/10.1007/978-3-642-18123-8_13} {\path{doi:10.1007/978-3-642-18123-8_13}}.

\bibitem{butterfly}
M.~García-Quismondo, J.~M. Reed, F.~S. Chew, M.~A.~M. del Amor, M.~J. Pérez-Jiménez, Evolutionary response of a native butterfly to concurrent plant invasions: Simulation of population dynamics, Ecological Modelling 360 (2017) 410--424.
\newblock \href {https://doi.org/https://doi.org/10.1016/j.ecolmodel.2017.06.030} {\path{doi:https://doi.org/10.1016/j.ecolmodel.2017.06.030}}.

\bibitem{Paun98}
{\relax Gh}.~P\u{a}un, Computing with membranes, Tech. Rep. 208, Turku Centre for Computer Science, Turku, Finland (November 1998).

\bibitem{10.1145/3431234}
B.~Song, K.~Li, D.~Orellana-Martín, M.~J. Pérez-Jiménez, I.~PéRez-Hurtado, \href{https://doi.org/10.1145/3431234}{A survey of nature-inspired computing: Membrane computing}, ACM Comput. Surv. 54~(1) (Feb 2021).
\newblock \href {https://doi.org/10.1145/3431234} {\path{doi:10.1145/3431234}}.
\newline\urlprefix\url{https://doi.org/10.1145/3431234}

\bibitem{DBLP:journals/nc/CardonaCMPPPS11}
M.~Cardona, M.~{\`{A}}. Colomer, A.~Margalida, A.~Palau, I.~Pérez{-}Hurtado, M.~J. Pérez{-}Jiménez, D.~Sanuy, A computational modeling for real ecosystems based on {P} systems, Nat. Comput. 10~(1) (2011) 39--53.
\newblock \href {https://doi.org/10.1007/S11047-010-9191-3} {\path{doi:10.1007/S11047-010-9191-3}}.

\bibitem{DBLP:journals/isci/Garcia-Victoria22}
P.~García{-}Victoria, M.~Cavaliere, M.~A. Gutiérrez{-}Naranjo, M.~Cárdenas{-}Montes, Evolutionary game theory in a cell: {A} membrane computing approach, Inf. Sci. 589 (2022) 580--594.
\newblock \href {https://doi.org/10.1016/J.INS.2021.12.109} {\path{doi:10.1016/J.INS.2021.12.109}}.

\bibitem{populationgames}
B.~Shulman, Population games and evolutionary dynamics, American Mathematical Monthly 119 (04 2012).
\newblock \href {https://doi.org/10.4169/amer.math.monthly.119.04.352} {\path{doi:10.4169/amer.math.monthly.119.04.352}}.

\bibitem{BrownNeumann+1951+73+80}
G.~W. Brown, J.~von Neumann, 6. SOLUTIONS OF GAMES BY DIFFERENTIAL EQUATIONS, Princeton University Press, Princeton, 1951, pp. 73--80.
\newblock \href {https://doi.org/doi:10.1515/9781400881727-007} {\path{doi:doi:10.1515/9781400881727-007}}.

\bibitem{DBLP:conf/eucc/Martinez-Piazuelo22}
J.~Martinez{-}Piazuelo, C.~Ocampo{-}Martinez, N.~Quijano, Generalized nash equilibrium seeking in population games under the brown-von neumann-nash dynamics, in: European Control Conference, {ECC} 2022, London, United Kingdom, July 12-15, 2022, {IEEE}, 2022, pp. 2161--2166.
\newblock \href {https://doi.org/10.23919/ECC55457.2022.9838437} {\path{doi:10.23919/ECC55457.2022.9838437}}.

\bibitem{PAUN2000108}
G.~Păun, Computing with membranes, Journal of Computer and System Sciences 61~(1) (2000) 108--143.
\newblock \href {https://doi.org/https://doi.org/10.1006/jcss.1999.1693} {\path{doi:https://doi.org/10.1006/jcss.1999.1693}}.

\bibitem{paunactive}
G.~P\u{a}un, P systems with active membranes: attacking np-complete problems, J. Autom. Lang. Comb. 6~(1) (2001) 75–90.

\bibitem{EulerMethodBook}
J.~Butcher, Numerical Methods for Ordinary Differential Equations, 2016.
\newblock \href {https://doi.org/10.1002/9781119121534} {\path{doi:10.1002/9781119121534}}.

\bibitem{DBLP:journals/nc/Gutierrez-NaranjoPR08}
M.~A. Gutiérrez{-}Naranjo, M.~J. Pérez{-}Jiménez, D.~Ramírez{-}Martínez, A software tool for verification of spiking neural {P} systems, Nat. Comput. 7~(4) (2008) 485--497.
\newblock \href {https://doi.org/10.1007/S11047-008-9083-Y} {\path{doi:10.1007/S11047-008-9083-Y}}.

\bibitem{MecosimRef16}
M.~García-Quismondo, R.~Gutiérrez-Escudero, I.~Pérez-Hurtado, M.~J. Pérez-Jiménez, A.~Riscos-Núñez, An overview of p-lingua 2.0, in: G.~P{\u{a}}un, M.~J. Pérez-Jiménez, A.~Riscos-Núñez, G.~Rozenberg, A.~Salomaa (Eds.), Membrane Computing, Springer Berlin Heidelberg, Berlin, Heidelberg, 2010, pp. 264--288.

\bibitem{MecosimRef17}
I.~Pérez-Hurtado, L.~Valencia-Cabrera, M.~Pérez-Jiménez, M.~Colomer, A.~Riscos-Núñez, Mecosim: A general purpose software tool for simulating biological phenomena by means of p systems, in: 2010 IEEE Fifth International Conference on Bio-Inspired Computing: Theories and Applications (BIC-TA), 2010, pp. 637--643.
\newblock \href {https://doi.org/10.1109/BICTA.2010.5645199} {\path{doi:10.1109/BICTA.2010.5645199}}.

\bibitem{UPSimulator}
P.~Guo, C.~Quan, L.~Ye, Upsimulator: A general p system simulator, Knowledge-Based Systems 170 (2019) 20--25.
\newblock \href {https://doi.org/https://doi.org/10.1016/j.knosys.2019.01.013} {\path{doi:https://doi.org/10.1016/j.knosys.2019.01.013}}.

\bibitem{UPSimulator2}
P.~Guo, C.~Quan, L.~Ye, A simulator for cell-like p system, in: J.~Qiao, X.~Zhao, L.~Pan, X.~Zuo, X.~Zhang, Q.~Zhang, S.~Huang (Eds.), Bio-inspired Computing: Theories and Applications, Springer Singapore, Singapore, 2018, pp. 223--235.

\bibitem{Cameron-1994}
P.~J. Cameron, Combinatorics: Topics, Techniques, Algorithms, Cambridge University Press, 1994.

\end{thebibliography}
\end{document}